\documentclass[a4paper,fleqn,usenatbib]{mnras}

\usepackage[T1]{fontenc}
\usepackage{ae,aecompl}

\usepackage{graphicx}	
\usepackage{amsmath}	
\usepackage{amssymb}	

\title[High Precision Polarimeter]{A high precision polarimeter for small telescopes}

\author[Bailey et al.]{
Jeremy Bailey,$^{1,2}$\thanks{E-mail: j.bailey@unsw.edu.au}
Daniel V. Cotton,$^{1,2}$
Lucyna Kedziora-Chudczer$^{1,2}$
\\
$^{1}$School of Physics, UNSW Australia, NSW 2052, Australia\\
$^{2}$Australian Centre for Astrobiology, UNSW Australia, NSW 2052, Australia\\
}

\date{Accepted XXX. Received YYY; in original form ZZZ}

\pubyear{2015}

\begin{document}
\label{firstpage}
\pagerange{\pageref{firstpage}--\pageref{lastpage}}
\maketitle

\begin{abstract}
We describe Mini-HIPPI (Miniature HIgh Precision Polarimetric Instrument), a stellar polarimeter weighing
just 650 grams but capable of measuring linear polarization to $\sim10^{-5}$. Mini-HIPPI is 
based on the use of a Ferroelectric Liquid Crystal (FLC) modulator. It can easily be mounted on a small
telescope and allows study of the polarization of bright stars at levels of precision which are hitherto largely
unexplored. We present results obtained with Mini-HIPPI on a 35 cm telescope. Measurements of polarized
standard stars are in good agreement with predicted values. Meaurements of a number of bright stars agree well with
those from other high-sensitivity polarimeters. Observations of the binary system Spica show polarization variability
around the orbital cycle.
\end{abstract}




\section{Introduction}

Recently a number of polarimeters have been developed that can measure stellar polarization at
parts-per-million levels \citep{hough06,wiktorowicz08,wiktorowicz15a,bailey15}. All of these instruments
are on 3--5m class telescopes and one of their main aims has been the measurement of polarized reflected
light from extrasolar planets \citep[e.g.][]{lucas09,wiktorowicz15b,bott16}. However, such instruments
can also provide valuable data on the polarization of bright stars. For example, surveys of bright stars
\citep{bailey10,cotton16} have identified interstellar polarization in relatively nearby stars, and found
that intrinsic polarization in normal stars is more common than had previously been thought. For studies of 
the polarization of bright stars it is not necessary to use a telescope as large as
3--5m.

Most of these instruments have been based on the photoelastic modulator (PEM) technology originally developed by James Kemp 
\citep{kemp69,kemp81,kemp87}. However, \citet{bailey15} showed that high polarimetric sensitivity could also be
achieved using Ferroelectric Liquid Crystal (FLC) modulators. The compact size and simple drive requirements
of FLC's make them ideally suited to the development of extremely small and lightweight instuments.
Here we describe a FLC-based polarimeter sufficiently compact and lightweight to be mounted on even the
smallest telescopes.

\section{Instrument Description}

\subsection{Overview}

Mini-HIPPI is based on the same design principles as the HIPPI instrument \citep{bailey15} used on the
3.9m Anglo-Australian Telescope (AAT). HIPPI was commissioned in 2014 and has been used regularly on the AAT for a number of
applications \citep{bailey15,cotton16,bott16,marshall16}. Mini-HIPPI shares the same modulators, detectors, and data acquisition and data
reduction software.

\begin{figure}
\includegraphics[width=\columnwidth]{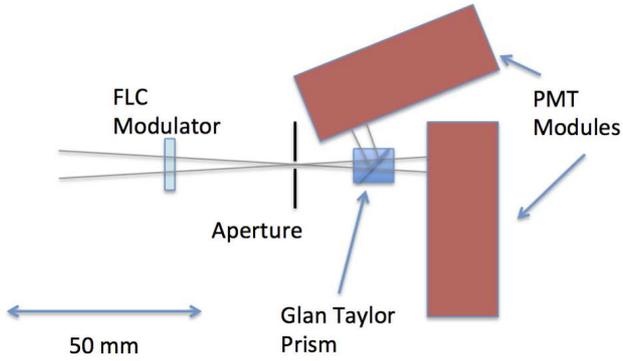}
\caption{Optical system of Mini-HIPPI}
\label{fig_mhippi1}
\end{figure}

\begin{figure}
\includegraphics[width=\columnwidth]{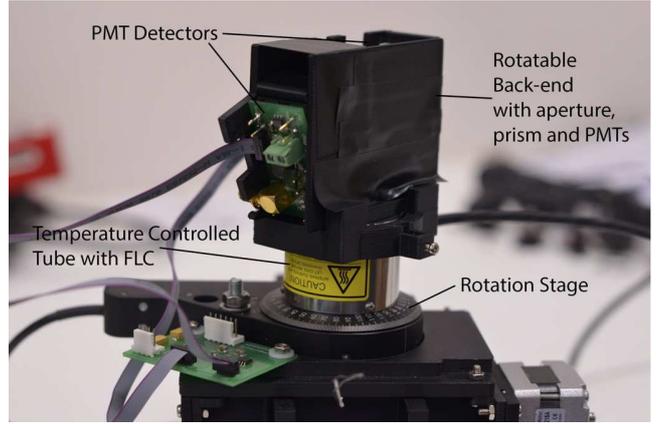}
\caption{Mini-HIPPI with the main components indicated}
\label{fig_mhippi3}
\end{figure}

The optical system of Mini-HIPPI is shown in figure \ref{fig_mhippi1} and a picture of the instrument is
given in figure \ref{fig_mhippi3}.

In order of their position on the light path from the telescope the components of the instrument are as
follows:

\begin{itemize}
\item A Thorlabs PRM1Z8 motorized rotation stage allows the whole instrument to be rotated.
\item The first optical element is the Ferroelectric Liquid Crystal (FLC) modulator. This is an
LV1300-AR-OEM device from Micron Technology. It is designed for the 400-700 nm wavelength range. It
operates as an electrically switched half-wave plate driven by a $\pm$5 V square wave which causes the
optical axis to switch between two orientations 45 degrees apart. The normal modulation frequency used is
500 Hz. Further details on these devices and the method of driving them can be found in \cite{bailey15}.
The FLC is a temperature sensitive device and is therefore mounted in a temperature controlled lens tube
(Thorlabs SM1L10H) and maintained at a temperature of 25$^\circ$C.
\item Behind the lens tube carrying the FLC, is a rotatable back-end module that carries the aperture,
polarizing prism and detectors. This can be adjusted in rotation to minimize the instrumental polarization as
described below. The aperture is mounted on a removable slide to allow change of aperture sizes.
\item The polarizing prism is a calcite Glan-Taylor prism (Thorlabs GT5-A) of 5mm aperture. This splits
the light into two beams with orthogonal polarizations, one passing straight through the prism, and one
directed upwards at an angle of 68 degrees to the incident beam.  
\item The two beams are fed to two Hamamatsu H10720-210 Photomultiplier Tube (PMT) modules with attached
transimpedance amplifiers, identical to those used with HIPPI \citep{bailey15}.
\end{itemize}

Apart from the commercial components described above, the construction of the instrument is entirely by 3D
printing in black ABS plastic.

The main differences from HIPPI are:

\begin{itemize}
\item The use of a Glan-Taylor prism rather than a Wollaston prism. This makes possible a much shorter
light path helping to keep the instrument compact.
\item The lack of any lenses. HIPPI used lenses to collimate the beam through the prism, and Fabry
lenses to image the pupil onto the detectors. The extremely short light path in Mini-HIPPI allows the
diverging beam from the aperture to be fed straight to the detectors while still being well within the 8
mm diameter of the photocathodes.
\item The lack of a motorized back-end rotation. This means that Mini-HIPPI only has two stages of
modulation (The primary FLC 500 Hz modulation, and the whole instrument rotation) whereas HIPPI has three.
\item Mini-HIPPI does not have a filter wheel. It has been designed to allow a fixed 25 mm diameter
filter to be fitted, but we normally use Mini-HIPPI without a filter, giving a broad bandpass defined mostly by the PMT
response. This helps to maximize the signal with small telescopes.
\end{itemize}

While all the observations described here use blue-sensitive PMTs and a modulator optimized for blue wavelengths, it is realatively
straightforward to adapt the instrument for red wavelengths by fitting a red-optimized FLC modulator and red-sensitive Hamamatsu
H10720-20 PMT modules with extended red multialkali photocathodes sensitive out to 920 nm.

\subsection{Data Acquisition System}

The data acquisition system for mini-HIPPI is similar to that for HIPPI and uses essentially the same
software developed in the National Instruments LabVIEW graphical programming environment. However, the
control electronics has been made more compact by using an Intel NUC miniature PC, running
Windows 7 Professional, and connected to two National Instruments USB-6211 Data Acquisition modules. These provide
analog inputs to read the signals from the PMTs and analog outputs to provide the drive waveform to the
FLC modulator as well as the control signals to set the PMT gain and HT voltage.

In operation the signals are read at a 10 microsecond sample time, synchronized with the modulation using
a trigger pulse derived from the FLC drive waveform. The signals are folded over the modulation period
(2 milliseconds for our standard 500 Hz modulation frequency) and written to output files after an
integration time of typically one second. The format of the output files is identical to those written by
HIPPI and can be analysed using the same data reduction software.

A simple quick-look data reduction system built into the acquisition software allows polarization levels
to be assessed in real time.

\begin{figure}
\includegraphics[width=\columnwidth]{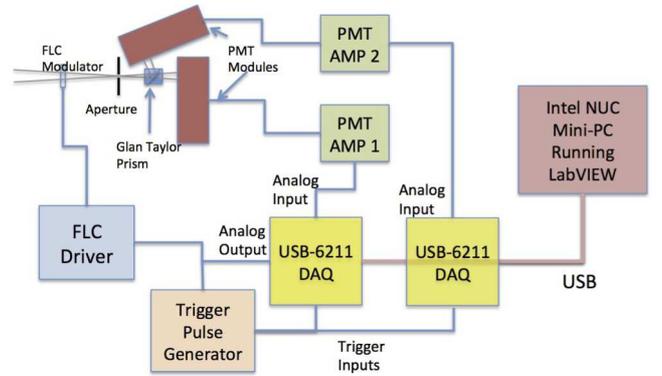}
\caption{Mini-HIPPI data acquistion and control system}
\label{fig_mhippi_control}
\end{figure}

\subsection{Observing Procedure} 

As explained by \citet{bailey15} polarimeters of this design inherently have a substantial instrumental
polarization due to multiple reflection effects in the FLC modulator \citep{gisler03,ovelar12}. The key to
achieving high precision is designing the instrument in a way that this instumental polarization can be
removed. This is possible because the instrument only measures one Stokes parameter at a time. By rotating
the back-end of the instrument relative to the FLC, a one-time adjustment when the instrument is set up, a
setting can be found at which the instrumental polarization is orthogonal to the Stokes parameter being
measured. This reduces the instrumental polarization to around 100 parts-per-million or less.

This residual instrumental polarization is then eliminated by repeating all observations with the whole
instrument rotated through 90 degrees relative to the telescope and sky. This reverses the measured sign of any
real polarization, while leaving the instrumental polarization unchanged. In practice we observe at four
position angles (0, 45, 90, 135 degrees). One Stokes parameter is obtained from the 0 and 90 degree
observations, and the other from the 45 and 135 observations.

This measured polarizations will still include any polarization introduced by the telescope (Telescope
Polarization or TP). This must be corrected by observations of low polarization stars as described later.

\subsection{Data Reduction}

The data reduction software used for Mini-HIPPI is identical to that for HIPPI described in
\citet{bailey15}. The polarization is determined by fitting a modelled curve to the observed modulation
waveforms using a Mueller matrix model of the instrument. The Mueller matrix for the instrument is the
matrix $\mathbf{M}$ that transforms the input Stokes vector of the source being observed $\mathbf{s_{in}}$ 
with components (I, Q, U, V) into the output Stokes vector $\mathbf{s_{out}}$ seen at the detector.

\begin{equation}
\label{eqn_mull}
\mathbf{s_{out}} = \mathbf{M} \mathbf{s_{in}}
\end{equation}

The matrix $\mathbf{M}$ is itself the product of the Mueller matrices for each of the optical elements in the system
as follows:

\begin{equation}
\mathbf{M} = \mathbf{M_{Pol} M_{Ret} M_{Depol}}
\end{equation}

where $\mathbf{M_{Pol}}$ describes the polarizing effect of the Wollaston prism, $\mathbf{M_{Ret}}$
describes the retardance of the modulator, and $\mathbf{M_{Depol}}$ describes the depolarizing effect of the
modulator. The matrix $\mathbf{M}$ varies around the modulation cycle because the retardance and
depolarization vary. We can determine the form of this variation by a laboratory calibration procedure where known polarization states are input to the instrument using a
lamp and rotatable polarizer. From the set of matrices $\mathbf{M}$ at different modulation phases we can 
then define a system matrix $\mathbf{W}$ which when multiplied by the input Stokes vector gives the observed
vector of output intensities at each modulation phase. From the inverse of this matrix \citep{sabatke00} we
can then derive the input polarization from the observed modulation curve. Full details of this procedure are given in \citet{bailey15}.

\begin{table}
\caption{Effective wavelength and modulation efficiency for different spectral types according to bandpass model}
\label{tab_sptype}
\begin{tabular}{ccc}
\hline Spectral & Effective  & Modulation  \\
 Type &  Wavelength (nm) &  Efficiency (\%) \\  \hline
B0 V  &  438.7  &   69.2 \\
A0 V  &  458.2  &   78.8 \\
F0 V  &  467.5  &   80.4 \\
G0 V  &  478.2  &   82.2 \\
K0 V  &  490.3  &   85.0 \\
M0 V  &  509.9  &   86.6 \\
M5 V  &  510.4  &   86.2 \\
\hline
\end{tabular}
\end{table}

\begin{figure}
\includegraphics[width=\columnwidth]{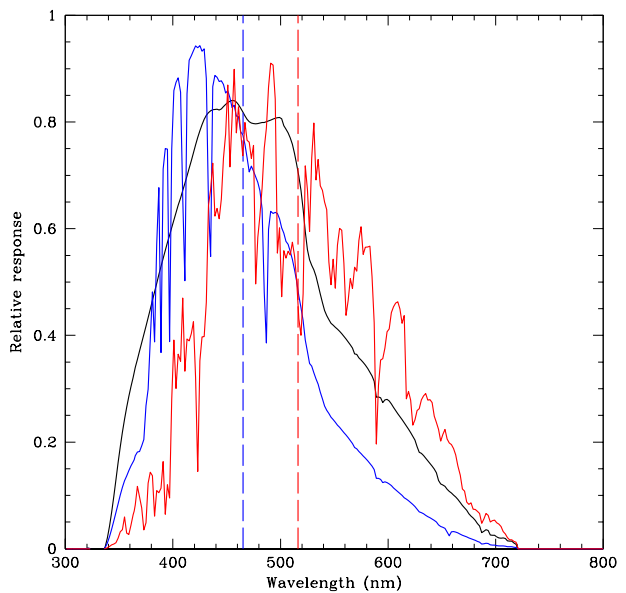}
\caption{The black line is the instrument response of Mini-HIPPI obtained by combining the atmospheric and
instrument throughput and the response of the PMT detector. The blue and red curves show the effective
response when observing an A0 star and an M0 star (using Kurucz model spectra). The dashed lines show the
effective wavelength in each case which shifts from 458.2 nm for the A0 to 509.9 nm for the M0 star. See also table \ref{tab_sptype}.}
\label{fig_mhippi_bandpass}
\end{figure}

\begin{figure}
\includegraphics[width=\columnwidth]{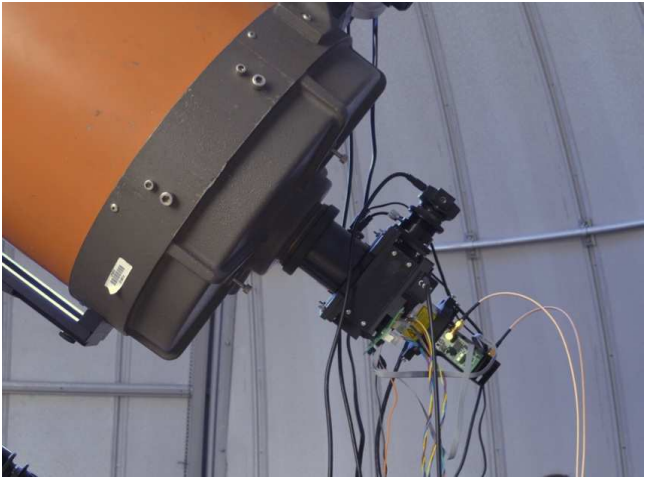}
\caption{Mini-HIPPI mounted on the 35cm telescope at UNSW Observatory}
\label{fig_telescope}
\end{figure}

\subsection{Bandpass Model}

Mini-HIPPI is normally used without a filter and therefore has a broad spectral response from around 350
-- 700 nm but with the peak in the blue at around 440 nm due to the response of the Ultra-bialkali
photocathodes used. This broad response means that the precise effective wavelength depends on the colour
of the object being observed. The modulation efficiency of the polarimeter is also wavelength dependent.
The FLC modulators used are designed to be half-wave plates at about 500 nm. The modulation efficiency
falls off at wavelengths either side of this. 

To account for these effects we use a bandpass model as described in \citet{bailey15}. This model allows
us to specify a source spectrum, usually one of the stellar atmosphere models from \cite{castelli04},
optionally correct it for interstellar extinction, and then apply atmospheric transmission and instrument
response functions to allow determination of the effective wavelength, and bandpass averaged modulation
efficiency. The typical results for nearby stars (no interstellar extinction) are given in table
\ref{tab_sptype}. Examples of response curves are given in figure \ref{fig_mhippi_bandpass}.

\section{Observations}

The observations described here were used to test and evaluate Mini-HIPPI and were made at the UNSW
Observatory located at the Kesington campus in Sydney. The telescope is a 35 cm Schmidt Cassegrain
(Celestron-14) mounted on a Losmandy German Equatorial mount. We attached Mini-HIPPI to the telescope via
an acquisition module that allowed the field-of-view to be imaged via a flip-in mirror, that can then be
removed to pass the starlight through to the instrument. The weight of the instrument plus acquisition
module is 1.2 kg. It attaches to the telescope through a standard 2-inch eyepiece fitting. The instrument
mounted on the telescope is shown in figure \ref{fig_telescope}. A camera
mounted on an attached 80 mm guide telescope is used for field acquisition and autoguiding.

The focal length of the telescope is 3900 mm and our 1.2 mm aperture corresponds to 63 arc seconds on the
sky. With this large aperture and our city site, which is never really dark even when the Moon is absent,
it is important to correct for sky background. Each observation consists of measurements at four rotator 
position angles, and at each of these angles we also make at least one sky observation. For long integrations
on fainter targets we take sky measurements before and after the target observations at each angle.
 The sky modulation curve is
subtracted from the target observation before further analysis.

\subsection{Telescope Polarization}

All observations need to be corrected for telescope polarization (TP). The TP is determined by
observations of nearby stars that we believe to have very low polarization. The low polarization
calibrators were all stars we have used previously for this purpose with HIPPI \citep{bailey15,cotton16}
or PlanetPol \citep{hough06, bailey10}. They are Sirius, $\beta$ Leo (BS 4534), $\beta$ Vir (BS 4540),
$\beta$ Hyi (HIP 2021, BS 98) and $\alpha$ Ser (BS 5854). Based on our previous observations of these
objects we believe they all have polarizations of $\sim$ 5 ppm (parts per million) or smaller.

\begin{table}
\caption{Low polarization star measurements to determine telescope polarization}
\begin{tabular}{l@{\hskip 8pt}l@{\hskip 8pt}l@{\hskip 8pt}l@{\hskip 8pt}l}
\hline Block & Star & Date & P (ppm) & $\theta$ (deg) \\  
\hline {\bf May 2016} & Sirius & May 13 & 95.2 $\pm$ 7.8 & 93.2 $\pm$ 2.3 \\
           & Sirius & May 14 & 98.8 $\pm$ 8.3 & 91.9 $\pm$ 2.5 \\
	   & Sirius & May 17 & 99.8 $\pm$ 8.2 & 90.6 $\pm$ 2.4 \\
	   & Sirius & May 19 & 75.8 $\pm$ 8.7 & 83.0 $\pm$ 3.2 \\
	 Average &   &       & {\bf 91.6 $\pm$ 4.1} & {\bf 90.1 $\pm$ 1.3 } \\
	\\
{\bf Jun 2016} & Sirius & Jun 9 & 62.1 $\pm$ 11.2 & 93.6 $\pm$ 5.5 \\
           & $\beta$ Leo & Jun 9 & 67.5 $\pm$ 19.7 & 89.2 $\pm$ 8.4 \\
	   & Sirius  & Jun 10 & 54.2 $\pm$ 6.6 & 93.3 $\pm$ 3.6 \\
	   & Sirius  & Jun 15 & 61.3 $\pm$ 12.5 & 91.9  $\pm$ 6.0 \\
	   & Sirius  & Jun 16 & 57.2 $\pm$ 7.6  & 91.2 $\pm$ 3.7  \\
	  Average &  &    &  {\bf 60.4 $\pm$ 5.6} & {\bf 91.8 $\pm$ 2.7} \\
	 \\
{\bf Jul-Sep} & $\beta$ Leo & Jun 29 & 31.7 $\pm$ 18.3 & 88.1 $\pm$ 16.2 \\
{\bf 2016}           & $\beta$ Leo & Jun 29 & 83.8 $\pm$ 18.4 & 82.0 $\pm$ 6.4 \\
	   & $\beta$ Vir & Jun 29 & 116.9 $\pm$ 29.5 & 95.6 $\pm$ 7.3 \\
	   & $\beta$ Leo  & Jul 1 & 64.4 $\pm$ 19.3 & 95.3 $\pm$ 8.2 \\
	   & $\beta$ Leo  & Jul 1 & 59.8 $\pm$ 19.7  & 93.6 $\pm$ 9.6  \\
	   & $\beta$ Hyi  & Jul 23 & 50.5 $\pm$ 26.7 & 93.0 $\pm$ 14.7 \\
	   & $\beta$ Hyi  & Jul 23 & 89.4 $ \pm$ 29.0 & 81.4 $\pm$ 19.3 \\
	   & $\beta$ Leo  & Jul 24 & 101.1 $ \pm$ 23.3 & 95.2 $\pm$ 6.5 \\
	   & $\beta$ Leo  & Jul 25 & 32.9 $\pm$ 22.0 & 88.9 $\pm$ 19.0  \\
	   & $\alpha$ Ser & Aug 11 & 89.4 $\pm$ 21.5 & 106.1 $\pm$ 6.8  \\
	   & $\beta$ Hyi & Aug 15  & 97.1 $\pm$ 19.3 & 108.5 $\pm$ 5.7  \\
	   & $\beta$ Hyi & Aug 31  & 77.8 $\pm$ 19.0 & 97.5 $\pm$ 7.0  \\
	   & Sirius      & Sep 19  & 59.3 $\pm$ 7.8  & 86.4 $\pm$ 4.1  \\
	   & Sirius      & Sep 19  & 83.0 $\pm$ 8.4  & 86.3 $\pm$ 3.2  \\
	 Average &  &   &  {\bf 71.0 $\pm$ 5.6 } & {\bf 93.5 $\pm$ 2.3} \\
\hline
\end{tabular}
\label{tab_tp}
\end{table}

We determined the TP separately for our blocks of observations in May 2016, Jun 2016 and Jul-Sep 2016. The
instrument was removed from the telescope and remounted and adjusted between each of these blocks. The TP
adopted was the mean of the polarization of all calibrators observed within each block, and this was used
to correct all other observations within the block.

When Sirius was observable this was our preferred calibrator as it could be observed to a precision of
usually better than 10 ppm in a short observation. The other objects were fainter and typical precisions
of individual observatons were $\sim$ 20 -- 30 ppm.

Between the May and June observing blocks the instrument was modified by fitting with an improved aperture
plate that removed some light leaks, and was remounted with a different orientation on the telescope. This
probably explains the change in TP between these runs.

\begin{table*}
\begin{center}
\caption{Predicted and observed polarization for polarized standard stars}
\begin{tabular}{llllllllllll}
\hline Star & Spectral Type & $E_{B-V}$ & $R_V$ & $P_{max}$ & $\lambda_{max}$ & K & $\theta$ (deg) &
\multicolumn{2}{c}{
$P$ (\%)} & $\theta$ (deg) & Refs \\
 & & & & & & & & Predicted & Observed & Observed & \\  \hline
HD 84810 & F8-K0Ib V & 0.34 & 3.1 & 1.62 & 0.57 & 1.15 & 100.0 & 1.566 & 1.570$\pm$.002 & 99.9$\pm$0.1 & 1,2 \\
HD 147084 & A4 II/III & 0.72 & 3.9 & 4.34 & 0.67 & 1.15 & 32.0 & 3.823 & 3.803$\pm$.005 & 31.8$\pm$0.1 & 3,4 \\
HD 154445 & B1V  & 0.42 & 3.15 & 3.73 & 0.558 & 0.95 & 90.1 & 3.576 & 3.629$\pm$.010 & 89.4$\pm$0.1 & 1,2,3 \\
HD 160529 & A2 Ia & 1.29 & 3.1 & 7.31 & 0.543 & 1.15 & 20.4 & 7.139 &  7.143$\pm$.022 & 20.8$\pm$0.1 &  1,2 \\
HD 187929 & F6-G0 I & 0.18 & 3.1 & 1.76 & 0.56 & 1.15 & 93.8 & 1.696 &  1.684$\pm$.006 & 92.9$\pm$0.1 &  1 \\
\hline
\end{tabular}

References:   1. \citet*{serkowski75}   2.  \citet{hsu82},  3.   \citet{wilking80},   
4.  \citet{martin99}  
\label{tab_stand}
\end{center}
\end{table*}

\subsection{High Polarization Stars}

We have observed a number of high polarization stars. These are used to calibrate the position angle of polarization and
also provide a check on the efficiency calibration of the instrument. The stars observed and the parameters needed for
our bandpass model are listed in table \ref{tab_stand}. 

These stars are distant stars polarized by the
interstellar medium. The wavelength dependence of interstellar polarization can be represented by the empirical model
\citep{serkowski75,wilking80}

\begin{equation}
P(\lambda) = P_{max} \exp \left( -K \ln^2 \frac{\lambda}{\lambda_{max}} \right)
\end{equation}
 
where the values of $P_{max}$, $\lambda_{max}$ and $K$ are empirical parameters that are fitted to
observed wavelength dependent polarization measurements and are given in table \ref{tab_stand} for our
standard stars. Since the wavelength dependence is well determined we can calculate the polarization we expect to observe
with Mini-HIPPI by integrating over the passband as described by \citet{bailey15}. The predicted polarizations are given
in table \ref{tab_stand}.

We observed HD 84810 on 7 occasions, HD 147084 twice, and the other stars once. The average results are given in table
\ref{tab_stand}. It can be seen that the agreement between predicted and observed polarization is excellent. The largest
discrepancy between observed and predicted is 0.05\% for HD 154445. Other objects agree to 0.01 -- 0.03\%. This is as
good as can be expected considering the accuracy of the past results we are comparing with. The results demonstrate the accuracy
of the instrument and of the bandpass model we are using to determine the efficiency corrections.

The position angles all agree with previous values to within one degree.

\begin{table*}
\begin{center}
\caption{Observations of low polarization stars compared with HIPPI and PlanetPol observations}
\begin{tabular}{llllllllll}
\hline Star & HD & Distance & N obs & \multicolumn{2}{c}{Mini-HIPPI (350-700 nm)} & \multicolumn{2}{c}{HIPPI (400-550nm)} & \multicolumn{2}{c}{PlanetPol (590-1000nm)} \\ 
 & & (pc) &  & $P$ (ppm) & $\theta$ (deg) & $P$ (ppm) & $\theta$ (deg) & $P$ (ppm) & $\theta$ (deg) \\  \hline
Sirius & 48915 & 2.6 & 11 & 3.1 $\pm$ 2.7 & 8 $\pm$ 25 &  5.5 $\pm$ 1.7 & 114 $\pm$ 18 \\
$\beta$ Hyi & 2151 & 7.5 & 4 & 14.8 $\pm$ 11.5 & 127 $\pm$ 23 & 8.8 $ \pm$ 2.5 & 95 $\pm$ 16 \\
$\beta$ Leo & 102647 & 11.1 & 7 & 11.8 $\pm$ 7.8 & 21 $\pm$ 19 & & & 2.3 $\pm$ 1.1 & 35 $\pm$ 14 \\
Fomalhaut & 216956 & 7.7 & 6 & 17.8 $\pm$ 5.9 & 73 $\pm$ 10 & 24.3 $\pm$ 3.1 & 111 $\pm$ 7 \\
Arcturus & 124897 & 11.3 & 6 & 17.9 $\pm$ 4.4 & 7 $\pm$ 7 & & & 6.3 $\pm$ 1.6 & 31 $\pm$ 7 \\
$\alpha$ Cen & 128620/1 & 1.3 & 26 & 15.1 $\pm$ 1.9 & 34 $\pm$ 4 \\
Canopus & 45348 & 94.8 & 6 & 122.8 $\pm$ 4.0 & 115 $\pm$ 1 & 112.8 $\pm$ 1.7 & 116 $\pm$ 1 \\
Acrux & 108248 & 98.7 & 1 & 327.9 $\pm$ 14.1 & 114 $\pm$ 1 & 363.2 $\pm$ 1.7 & 113 $\pm$ 1 \\
$\epsilon$ Sgr &  169022 & 43.9 & 2  & 112.7 $\pm$ 11.4 & 34 $\pm$ 3 & 115.3 $\pm$ 1.9 & 37 $\pm$ 1 \\
\hline
\end{tabular}
\label{tab_comp}
\end{center}
\end{table*} 

\subsection{Comparison with other high precision polarimeters}

In table \ref{tab_comp} we compare Mini-HIPPI observations of a number of stars, most observed on multiple
nights, with previous high-precision observations of these stars. Some of these earlier observations are
with HIPPI on the Anglo-Australian Telescope, either from \citet{cotton16}, or from new
observations first reported here in the case of $\epsilon$ Sgr and Acrux. Other observations are with
PlanetPol on the William Herschel Telescope \citep*{bailey10}. In making these comparisons it is important to note that
the wavelengths are not the same. The HIPPI observations are in a narrower band, the SDSS $g'$ band from about 400 -- 550
nm, and the PlanetPol observations are in a much redder band from 590 -- 1000 nm. The first three stars listed are stars 
that we use as low polarization calibrators, so it is not surprising that the measured polarizations are low. The next
three stars (Fomalhaut, Arcturus and $\alpha$ Cen) are nearby bright stars that we are evaluating as potential bright low
polarization calibrators. Canopus is a somewhat more distant object, that shows polarization, probably interstellar in
origin. Acrux and $\epsilon$ Sgr \citep{cotton16b} are B star systems that likely have both intrinsic and interstellar polarization contributions.
It can be seen that the agreement between the Mini-HIPPI observations and earlier observations is typically at
the $\sim$10 ppm level or better. Where there are somewhat larger differences these are entirely consistent with plausible effects
due to wavelength dependence.

At low levels of polarization where the polarization is of a similar magnitude to the error,
observed values of polarization are subject to a positive bias \citep{simmons85}. Polarization data are
sometimes corrected by applying a debiasing correction of which one common form is to replace $P$ by
$\sqrt{P^2-\sigma_P^2}$. Following our procedure in earlier publications \citep{bailey10,cotton16} we have
not applied any debiasing correction to the data tabulated here. We apply such corrections only when they
are needed for specific statistical analyses where the bias becomes important. We also note that in most
cases the problem caused by this bias can be avoided by working in normalized Stokes parameters ($Q/I$ and
$U/I$) rather than in $P$ and $\theta$.

It can be seen from the results in table \ref{tab_comp} that Mini-HIPPI can easily detect polarization
at levels $\sim$100 ppm as shown by the results on Canopus and $\epsilon$ Sgr, which agree very well in
polarization and position angle with previous observations. In principle detection of polarization levels
down to $\sim20-30$ ppm should be possible, but this is hard to test with the limited existing data on
polarization at these levels.

It should be noted that Arcturus is reported by \citet{kemp86} to be a polarimetric variable with an amplitude of about
50 ppm in the B band and a possible period of around 45 days. Our current observations are insufficient to confirm such variability.

Our observations of $\alpha$ Cen apply to the combined light of both $\alpha$ Cen A and B. Although $\alpha$ Cen was previously
observed by HIPPI \citep{cotton16} the observation is considered unreliable because $\alpha$ Cen B would have been close to
the edge of the aperture.

\begin{figure}
\includegraphics[width=\columnwidth]{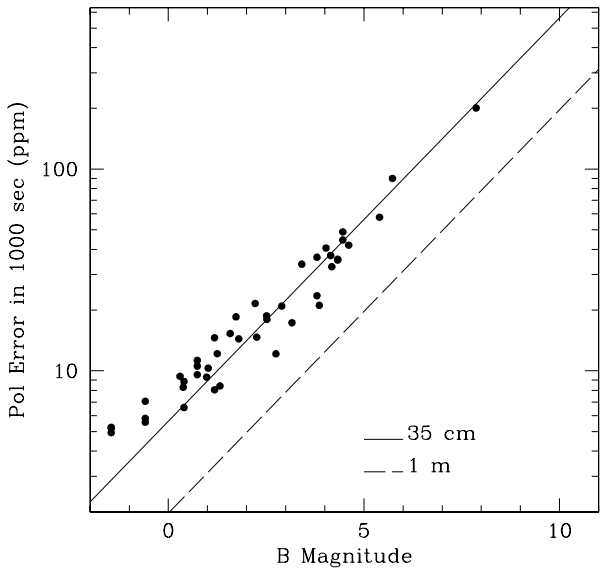}
\caption{Polarization errors of Mini-HIPPI observations scaled to 1000 seconds integration and plotted against
B-magnitude. The line has the slope expected for photon-noise limited performance and is fitted through the points for
the fainter stars. The dashed line is the expected performance for Mini-HIPPI on a 1 metre telescope.}
\label{fig_sn}
\end{figure}

\subsection{Performance versus Magnitude}

In figure \ref{fig_sn} we have plotted the polarization errors derived from the data reduction system for a range of
stars observed with Mini-HIPPI against the B magntitude of the stars. The measured errors have been scaled to a uniform
integration time ($T$) of 1000 sec under the assumption that the error scales as $T^{-0.5}$. The actual integration times
ranged from 400 to 1600 seconds. Over most of the range it can be seen that the error scales with magnitude in the way
expected for photon shot noise limited performance as shown by the line on the diagram. However, the performance is
relatively poorer (above the line) for the brightest stars. This could be an indication that the 500 Hz modulation we are
using is not fast enough to completely remove scintillation noise for the brightest stars. It may also be in part due to
the requirement to turn down the PMT gain by using a lower HT voltage to avoid saturation for these stars. The dashed
line on the diagram shows the performance expected for Mini-HIPPI on a 1 metre telescope.

\begin{figure}
\includegraphics[width=\columnwidth]{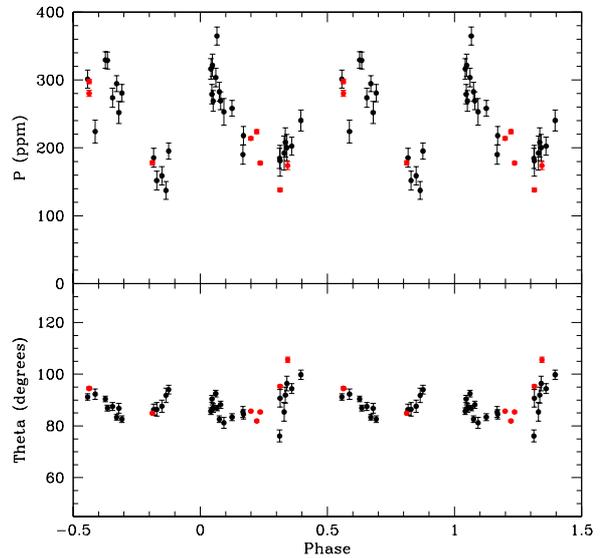}
\caption{Mini-HIPPI observations (in black) of the polarization and position angle of Spica plotted against phase on the orbital 
period. The red points are observations of Spica in the $g'$ band taken with HIPPI on the Anglo-Australian Telescope.}
\label{fig_spica}
\end{figure}

\subsection{Observations of Spica}

As an example of the type of science that can be done with Mini-HIPPI we present observations of Spica ($\alpha$ Vir).
Spica is a binary system consisting of two B stars orbiting with a period of just over 4 days. Previous polarimetry of
Spica was presented by \citet{pfeiffer77} who described it as unpolarized and gave a value of P = 0.03$\pm$ 0.01 \%.
\citet{tinbergen82} reported observations showing a marginal detection of polarization in the Q Stokes parameter.
\citet{elias08} also report a small polarization in an average of 11 measurements but say that the standard deviation
indicates that this star is not a polarization variable. \citet{cotton16} presented three measurements of Spica with 
HIPPI on the AAT that showed clear polarization and significant variability.

In figure \ref{fig_spica} we show observations of Spica obtained with Mini-HIPPI that show clear periodic variability in
polarization and position angle. The observations are plotted against phase on the oprbital cycle according to the
ephemeris of \citet{tkachenko16}. Zero phase corresponds to the time of periastron passage.

Also plotted on figure \ref{fig_spica} are eight observations of Spica obtained with HIPPI on the AAT in the $g'$ band.
Although this is a different band to the one used for Mini-HIPPI the effective wavelengths are similar. It can be seen
that the HIPPI and Mini-HIPPI observations appear quite consistent.

These results show that Mini-HIPPI, even on a very small telescope, is capable of easily detecting polarization variation, where
that was not detectable with a previous generation of polarimeters. 

We are continuing to monitor the polarization of Spica with the aim of filling in remaining gaps in the phase curve, and
we will present a full analysis at a later date. Spica is an important system, because it enables the polarimetric
modelling of a binary system to be compared with the results of interferometric observations \citep{herbison71}.

\section{Conclusions}

We have described the Mini-HIPPI polarimeter, a compact and lightweight instrument based on the use of a ferroelectric
liquid crystal modulator. Test observations with Mini-HIPPI on a 35 cm telescope show polarizations for standard stars in
excellent agreement with predicted values. We measure the telescope polarization to be $\sim$60 -- 90 ppm. A number of
stars observed with Mini-HIPPI show polarization results typically within about 10 ppm or better of previous measurements
with other high precision polarimeters. We have used Mini-HIPPI to study the polarization of the binary star system
Spica, and it easily detects periodic polarization variations around the binary cycle which were not detectable with the
precision available on earlier instruments.

\section*{Acknowledgements}

This work was supported by the Australian Research Council through Discovery Projects grant DP 160103231.









\bsp	
\label{lastpage}
\end{document}